\documentclass[a4paper,11pt]{article}

\usepackage{jcappub} 

\usepackage{chapterbib}    
\usepackage{color}         
\usepackage{graphics}      
\usepackage[T1]{fontenc} 

\title{\boldmath Axion searches with the EDELWEISS-II experiment}

\author[a]{E. Armengaud,}
\author[b]{Q. Arnaud,}
\author[b]{C. Augier,}
\author[c]{A. Benoit,}
\author[b]{A. Benoit,}
\author[d]{L. Berg\'e,}
\author[e]{T. Bergmann,}
\author[f,g]{J. Bl$\mbox{\"u}$mer,}
\author[d]{A. Broniatowski,}
\author[h]{V. Brudanin,}
\author[c]{P. Camus,}
\author[b]{A. Cazes,}
\author[b]{B. Censier,}
\author[d]{M.~Chapellier,}
\author[b]{F.~Charlieux,}
\author[d]{F.~Cou\"{e}do,}
\author[i]{P.~Coulter,}
\author[f]{G.A.~Cox,}
\author[a]{T.~de Boissière,}
\author[b]{M.~De Jesus,}
\author[d]{Y.~Dolgorouky,}
\author[d]{A.A.~Drillien,}
\author[d]{L.~Dumoulin,}
\author[g]{K.~Eitel,}
\author[h]{D.~Filosofov,}
\author[a]{N.~Fourches,}
\author[b]{J.~Gascon,}
\author[a]{G.~Gerbier,}
\author[a]{M.~Gros,}
\author[g]{L.~Hehn,}
\author[i]{S.~Henry,}
\author[a]{S.~Herv\'e,}
\author[f]{G.~Heuermann,}
\author[d]{N.~Holtzer,}
\author[d]{V.~Humbert,} 
\author[b]{A.~Juillard,}
\author[b,f]{C.~K\'ef\'elian,}
\author[e]{M.~Kleifges,}
\author[f]{H.~Kluck,}
\author[g]{V.~Kozlov,}
\author[i]{H.~Kraus,}
\author[j]{V.A.~Kudryavtsev,}
\author[d]{H.~Le Sueur,}
\author[d]{M.~Mancuso,}
\author[d]{C.~Marrache-Kikuchi,}
\author[d]{S.~Marnieros,}
\author[e]{A.~Menshikov,}
\author[a]{X-F.~Navick,}
\author[a]{C.~Nones,}
\author[d]{E.~Olivieri,}
\author[k]{P.~Pari,}
\author[a]{B.~Paul,}
\author[a,d]{M.C.~Piro,}
\author[d]{O.~Rigaut,}
\author[j]{M. Robinson,}
\author[h]{S.~Rozov,}
\author[b]{V.~Sanglard,}
\author[f]{B.~Schmidt,}
\author[g]{B.~Siebenborn,}
\author[e]{D.~Tcherniakhovski,}
\author[d]{M.~Tenconi,}
\author[b]{L.~Vagneron,}
\author[g]{R.J.~Walker,}
\author[e]{M.~Weber,}
\author[h]{E.~Yakushev,}
\author[i]{X.~Zhang}

\author{(The~EDELWEISS~Collaboration)}

\affiliation[a]{\small CEA, Centre d'Etudes Saclay, IRFU, 91191 Gif-Sur-Yvette Cedex, France}
\affiliation[b]{\small IPNL, Universit\'{e} de Lyon, Universit\'{e} Lyon 1, CNRS/IN2P3, 4 rue E. Fermi 69622 Villeurbanne cedex, France}
\affiliation[c]{\small CNRS-N\'{e}el, 25 Avenue des Martyrs, 38042 Grenoble cedex 9, France}
\affiliation[d]{\small CSNSM, Universit\'e Paris-Sud, IN2P3-CNRS, bat 108, 91405 Orsay,  France}
\affiliation[e]{\small Karlsruhe Institute of Technology, Institut f\"ur Prozessdatenverarbeitung und Elektronik, 76021 Karlsruhe, Germany}
\affiliation[f]{\small Karlsruhe Institute of Technology, Institut f\"ur Experimentelle Kernphysik, 76128 Karlsruhe, Germany}
\affiliation[g]{\small Karlsruhe Institute of Technology, Institut f\"ur Kernphysik, 76021 Karlsruhe, Germany}

\affiliation[h]{\small Laboratory of Nuclear Problems, JINR, Joliot-Curie 6, 141980 Dubna, Moscow region, Russia}
\affiliation[i]{\small University of Oxford, Department of Physics, Keble Road, Oxford OX1 3RH, UK}
\affiliation[j]{\small Department of Physics and Astronomy, University of Sheffield, Hounsfield Road, Sheffield S3 7RH, UK}	
\affiliation[k]{\small CEA, Centre d'Etudes Saclay, IRAMIS, 91191 Gif-Sur-Yvette Cedex, France}

\emailAdd{claudia.nones@cea.fr}
\emailAdd{thibault.main-de-boissiere@cea.fr}
\emailAdd{eric.armengaud@cea.fr}

\abstract{We present new constraints on the couplings of axions and more generic axion-like particles using data from the EDELWEISS-II experiment. The EDELWEISS experiment, located at the Underground Laboratory of Modane, primarily aims at the direct detection of WIMPs using germanium bolometers. It is also sensitive to the low-energy electron recoils that would be induced by solar or dark matter axions.
Using a total exposure of up to 448~kg.d, we searched for axion-induced electron recoils down to 2.5~keV within four scenarios involving different hypotheses on the origin and couplings of axions. We set a 95~\% CL limit on the coupling to photons $g_{A\gamma}<2.13\times 10^{-9}$~GeV$^{-1}$ in a mass range not fully covered by axion helioscopes. We also constrain the coupling to electrons, $g_{Ae} < 2.56\times 10^{-11}$, similar to the more indirect solar neutrino bound. Finally we place a limit on $g_{Ae}\times g_{AN}^{\rm eff}<4.70 \times 10^{-17}$, where $g_{AN}^{\rm eff}$ is the effective axion-nucleon coupling for $^{57}$Fe. Combining these results we fully exclude the mass range $0.91\,{\rm eV}<m_A<80$~keV  for DFSZ axions and $5.73\,{\rm eV}<m_A<40$~keV for KSVZ axions.}

\begin{document}
\maketitle
\flushbottom

\section{Introduction}
\label{sec:intro}

Following Peccei and Quinn's original solution~\cite{bib:pecceiandquinn} to the CP problem in QCD, Weinberg~\cite{bib:weinberg} and Wilczek~\cite{bib:wilczek} deduced the existence of a new, elusive pseudo-scalar particle, the axion. Both the axion mass and the strength of its couplings to ordinary particles are inversely proportional to the Peccei-Quinn symmetry-breaking scale $f_A$. While the original axion model with $f_A$ associated with the electroweak scale was quickly dismissed by subsequent experiments, "invisible" axions with $f_A$ as a free parameter are still viable. The most frequently studied are the so-called hadronic models such as KSVZ (Kim-Shifman-Vainstein-Zakharov)~\cite{bib:KSVZ} and the GUT models such as the DFSZ (Dine-Fischler-Srednicki-Zhitnitskii) model~\cite{bib:DFSZ}. These models still provide a solution to the strong CP problem. In both cases, the axion mass, $m_A$, is related to $f_A$:

\begin{equation}
m_{A}=\left[\frac{z}{(1+z+w)(1+z)}\right]^{\frac{1}{2}}\frac{f_{\pi}m_{\pi}}{f_{A}}=6\,{\rm eV} \times\left(\frac{10^{6}\, {\rm GeV}}{f_A}\right)
\label{eq:eq2}
\end{equation}

\noindent where $m_\pi=135$~MeV is the pion mass, $f_\pi \approx 92$~MeV the pion decay constant, while $z=m_u/m_d=0.56$ and $w=m_u/m_s=0.029$ are the mass ratios of the lightest up, down and strange quarks, with significant uncertainties especially for $z$~\cite{bib:PDG}. We use $\hbar=c=1$.

The effective axion couplings to photons ($g_{A\gamma}$), electrons ($g_{Ae}$) and nucleons ($g_{AN}$) are model dependent~\cite{bib:kaplan,bib:srednicki}. For example, hadronic axions are coupled to new, heavy quarks and do not interact with ordinary quarks and leptons at the tree level leading to a strong suppression of g$_{Ae}$. On the contrary, DFSZ axions require that Standard Model quarks and leptons carry a Peccei-Quinn charge. Experimental searches and astrophysical constraints can be translated to limits on $f_A$, or equivalently on the axion mass, within a given axion model.
On the other hand, axion-like particles (ALPs) are pseudo-scalar fields generically predicted by string theory~\cite{bib:string}, as the Kaluza-Klein zero modes of anti-symmetric tensor fields. The mass and couplings of ALPs are not directly related to their Peccei-Quinn-like scale. It is therefore relevant to also search for ALP couplings in a model-independent way.

The purpose of this article is to report on a study of the interaction of axions or ALPs produced by different mechanisms in the Sun or constituting the galactic dark matter halo (Section~\ref{sec:sources}) with EDELWEISS-II germanium bolometers operated underground and described in Section~\ref{sec:experiment}. Two detection mechanisms in the germanium detectors, described in Section~\ref{sec:detection}, are exploited: the coherent Bragg diffraction, related to $g_{A\gamma}$ and the axio-electric effect, which is the analogue of a photo-electric effect with the absorption of an axion instead of a photon. We benefit from the quality of the EDELWEISS-II data set: a large exposure corresponding to 14-month data taking with ten 400-g germanium detectors, a good energy resolution and a very low background down to an energy threshold of 2.5~keV. This low background was made possible thanks to both the low-radioactivity underground setup of the experiment and the so-called ID detector design~\cite{id_paper}  which allows a selection of interactions that takes place within a fiducial volume for each detector.
We carry out four axion searches associated with different benchmark scenarios. These searches are described in Sections~\ref{sec:primakoff}-\ref{sec:dm}. The derived constraints on axion couplings will be discussed, and in particular we interpret them within the specific DFSZ and KSVZ axion models.

\section{Possible axion sources: the Sun and the galactic halo}
\label{sec:sources}

The Sun could be a major source of axions. The different mechanisms which can lead to axion production in the Sun will be briefly reviewed in Section~\ref{sec:axion_sun}.  A second possibility is that axions constitute a major fraction of dark matter and are present in the galactic halo, as described in Section~\ref{sec:dm_search}.

\subsection{Axion production in the Sun}
\label{sec:axion_sun}
Several production mechanisms will be briefly described and considered later for data analysis:

\begin{enumerate}
  \item Primakoff production:
  $\gamma$  $\rightarrow$ A in the presence of charged particles
  \item Nuclear magnetic transition of $^{57}$Fe nuclei:
 $^{57}$Fe$^*$ $\rightarrow ^{57}$Fe +  A
  \item Compton-like scattering:
 e$^- + \gamma$ $\rightarrow$ e$^-$ + A
  \item Axion bremsstrahlung:
 e$^- \rightarrow$ e$^-$  + A in the presence of charged particles
 \item Axio-recombination:
 e$^-$ + I $\rightarrow$ I$^-$ + A where I is an ion
 \item Axio-deexcitation:
 I$^*$ $\rightarrow$  I + A where I$^*$ is an excited state of I
\end{enumerate}

\begin{figure}[!ht]
\centering
\includegraphics[width=\textwidth]{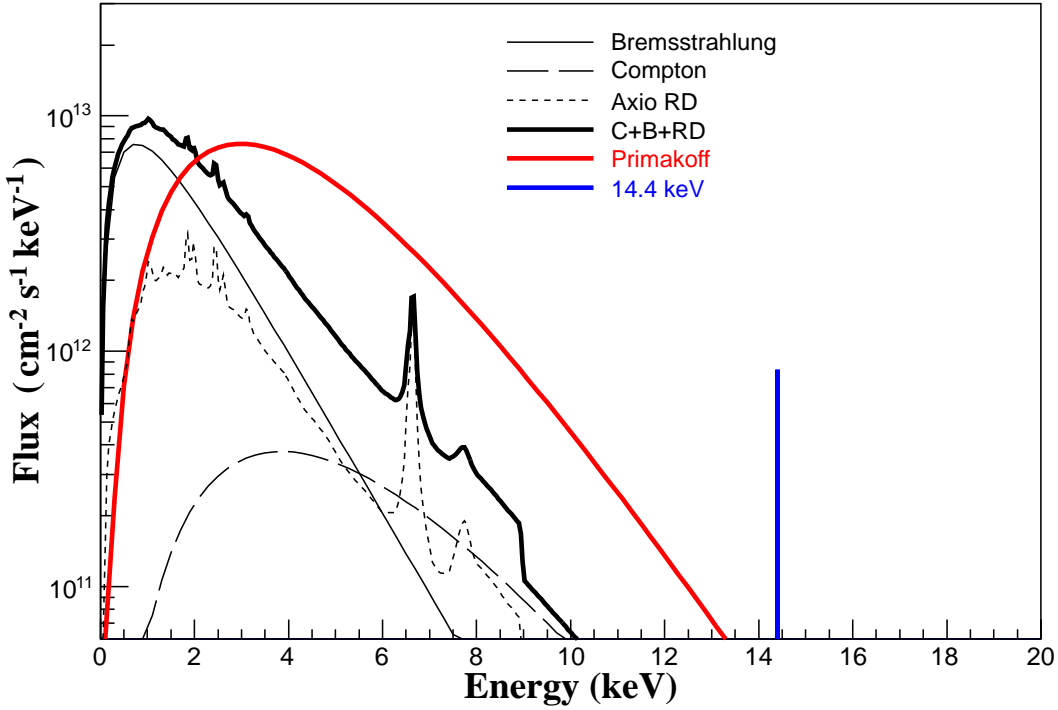}
\caption{Predicted solar axion fluxes in the EDELWEISS detectors from different mechanisms. The thick solid black line corresponds to the sum of Compton, bremsstrahlung and axio-RD (recombination-deexcitation). Red: Primakoff axions. Blue: $^{57}$Fe nuclear transition. The intrinsic width of this line, dominated by Doppler broadening, is 5 eV. The effective axion couplings corresponding to the represented fluxes are $g_{A\gamma}=10^{-9}$~GeV$^{-1}$, $g_{Ae}=10^{-11}$ and $g_{AN}^{\rm eff}=10^{-7}$.}
\label{fluxes}
\end{figure}

\noindent We will refer to the sum of axio-recombination and axio-deexcitation as the axio-RD mechanism. The relative intensity of the mechanisms is model-dependent. For example, in the case of non hadronic axions such as those described by the DFSZ model, fluxes related to Compton and bremsstrahlung processes are far more intense than those predicted by hadronic models for the same value of $f_A$. This is due to the fact that the coupling to electrons arises at the tree level. In this case, the Compton and bremsstrahlung channels for axion production largely prevail over the Primakoff effect (see Section~\ref{sec:primakoff}). On the contrary, the latter dominates hadronic axion emission. As for the $^{57}$Fe axions, whose flux depends only on the isoscalar and isovector coupling constants, the axion production rate is similar in hadronic and non hadronic models.
Fig.~\ref{fluxes} shows the evaluated fluxes on Earth for the various processes.

\subsubsection{Production by Primakoff effect}
\label{sec:primakoff_prod}

Axions could be efficiently produced in the Sun by the inverse Primakoff conversion of thermal photons in the electromagnetic field of the solar plasma. The effective Lagrangian of the axion-photon coupling is given by:

\begin{equation}
\mathcal{L}= -\frac{1}{4}g_{A\gamma}F^{\mu\nu}\tilde{F}_{\mu\nu}\,\phi_A=g_{A\gamma}\bf{E}\cdot\bf{B}\, \phi_A ,
\label{eq:lagr_gamma}
\end{equation}

\noindent where $F^{\mu\nu}$ is the electromagnetic field tensor,  $\tilde{F}_{\mu\nu}$ its dual, $\phi_A$ the axion field and g$_{A\gamma}$ the axion-photon effective coupling constant.

Within standard axion models, the  coupling  g$_{A\gamma}$, which has the dimension of (energy)$^{-1}$, can be written as:
\begin{equation}
\label{eq:gag}
g_{A\gamma}= \frac{\alpha}{2\pi f_A}\left[\frac{E}{N} - \frac{2(4+z+w)}{3(1+z+w)}\right]  
\end{equation}

\noindent where $\alpha$ is the fine structure constant and $E/N$ is the ratio of the electromagnetic to color anomalies of the Peccei-Quinn symmetry ($E/N=8/3$ and $0$ for DFSZ and KSVZ models, respectively).

The expected solar Primakoff axion flux was estimated in~\cite{bib:CAST_gay_2007}, and is well approximated by the expression where the energy $E$ is in keV:

\begin{equation}
\label{eq:solar_prim_flux}
\frac{d\Phi}{d{\text E}}=\frac{6.02\times 10^{14}}{{\rm cm}^2 \,{\rm keV}\,{\rm s}}\left(\frac{g_{A\gamma}\times 10^8}{{\rm GeV}^{-1}}\right)^2\hspace{1mm}{\text E}^{2.481}e^{-E/1.205}
\end{equation}

This corresponds to a broad spectrum, with an average energy of about 4.2 keV and a negligible intensity above 10 keV as shown in Fig.~\ref{fluxes}. The intensity of the resulting axion flux scales as g$_{A\gamma}^2$. Finally, we mention that this flux estimation is valid for ultra-relativistic axions. In~\cite{bib:dilella}, a calculation of the Primakoff flux is also given for non-relativistic axions. The derived correction to the flux in the mildly relativistic regime is of the order of 1~\% for $m_A=200$~eV.

\subsubsection{Production via $^{57}$Fe nuclear magnetic transition}
\label{sec:nucl_trans}

As axions couple to nucleons in many models, another possible mechanism of axion production in the Sun can be their emission following the de-excitation of the low-lying energy levels of some nuclei populated by the high solar temperature. The ability to measure keV-scale energy depositions with EDELWEISS bolometers is an incentive to study more specifically the 14.4~keV monochromatic axions emitted in the M1 transition of $^{57}$Fe nuclei. 

This particular isotope is considered because of its stability and its remarkable abundance among heavy elements in the Sun (the average  $^{57}$Fe density in the Sun's core is about $9 \times 10^{19}$ cm$^{-3}$~\cite{serenelli}). Last, but not least, its first excited nuclear state, placed at $E^{*}=14.4$~keV above the ground state, is low enough to be thermally excited in the hot interior of the Sun, where the average temperature is $kT\sim 1.3$~keV~\cite{Moriyama,haxtonlee}. The conventional relaxation of the excited $^{57}$Fe nucleus occurs through the emission of a 14.4 keV photon or  an internal-conversion electron. Since this de-excitation corresponds mainly to an M1 transition (E2/M1 mixing ratio is $0.002$), also an axion could be emitted.

The effective Lagrangian coupling axions to nucleons is given by:
\begin{equation} 
\label{eq:axion_nucleon_L}
\mathcal{L}=i \bar{\psi}_{N}\gamma_{5}(g_{AN}^{0}+g_{AN}^{3}\tau_{3}) \psi_{N} \phi_A
\end{equation}

\noindent where $\phi_A$ is the axion field, $\psi_{N}$ is the nucleon isospin doublet, and $\tau_{3}$ the associated isospin Pauli matrix. The two dimensionless parameters $g_{AN}^{0}$ and  $g_{AN}^{3}$ are the model-dependent isoscalar and isovector axion-nucleon coupling constants, respectively. In KSVZ models they are related to the scale $f_A$ by the following expressions~\cite{bib:kaplan,bib:srednicki}:

\begin{equation} \label{eq:ax_nucl_K}
\begin{split}
g^0_{AN} & =-7.8\times10^{-8} \left( \frac{6.2\times10^{6}\text{GeV}}{f_{A}} \right) \left( \frac{3F-D+2S}{3}\right) \\
g^3_{AN} & =-7.8\times10^{-8} \left( \frac{6.2\times10^{6}\text{GeV}}{f_{A}} \right) \left[(D+F)\frac{1-z}{1+z}\right]
\end{split}
\end{equation}

\noindent Here, the dimensionless constants $F=0.462$ and $D=0.808$~\cite{F_D_parameters} are invariant matrix elements of the axial current, determined by the hyperon semileptonic decays and flavor $SU(3)$ symmetry. The flavor-singlet axial-vector matrix element $S$ is still a poorly constrained dimensionless parameter. It can be estimated by measurements of the polarized nucleon structure function, but suffers from large uncertainties and ambiguity. Intervals for $S$ proposed in the literature lie in the range $0.15 - 0.55$~\cite{SpinMuon, Altarelli}. In the model-dependent analysis presented later, we will use the benchmark value $S=0.5$. 

In non-hadronic axions as in the DFSZ model, the values for $g_{\rm{AN}}^{0}$ and for $g_{\rm{AN}}^{3}$ depend on two additional unknown parameters, $X_u$ and $X_d$~\cite{bib:kaplan}. They are related to $\tan \beta _{DFSZ}$, the ratio of two Higgs vacuum expectation values of the model, by the relations $X_u+X_d=1$ and $X_d=\cos^2 \beta _{DFSZ}$.  The expressions for $g_{\rm{AN}}^{0}$ and for $g_{\rm{AN}}^{3}$ are given in this case by~\cite{bib:kaplan}:

\begin{equation} \label{eq:ax_nucl_D}
\begin{split}
g^0_{AN} & =5.2\times10^{-8} \left( \frac{6.2\times10^{6}\,\text{GeV}}{f_{A}} \right)
\left[ \frac{(3F-D)(X_u-X_d-3)}{6}+\frac{S(X_u+2X_d-3)}{3} \right]
\\
g^3_{AN} & =5.2\times10^{-8} \left( \frac{6.2\times10^{6}\,\text{GeV}}{f_{A}} \right)
\frac{D+F}{2} \left( X_u-X_d-3 \frac{1-z}{1+z} \right) \ .
\end{split}
\end{equation}

In the later model-dependent studies, we will take $\cos^2\beta _{DFSZ}=1$ as a benchmark value. Note that this choice also maximises the axio-electric cross section in the DFSZ model, as we will see in Section~\ref{sec:compton}.

We discuss now in detail the decay of the 14.4~keV first excited state of the $^{57}$Fe nucleus to the ground state via axion emission, a process that competes with ordinary M1 and E2 gamma transitions. In general, the axion-to-photon emission rate ratio for the M1 nuclear transition calculated in the long-wavelength limit is~\cite{frank_88}:

\begin{eqnarray} \label{eq:axionphotonration}
\frac{\Gamma_{\rm{A}}}{\Gamma_{\gamma}} = \left(
\frac{k_{\rm{A}}}{k_{\gamma}}\right) ^{3} \: \frac{1}{2\pi\alpha}\: \frac{1}{1+\delta^{2}}\: \left[\frac{g_{\rm{AN}}^{0}\beta +
g_{\rm{AN}}^{3}}{(\mu_{0}-1/2)\beta + \mu_{3} - \eta}\right] ^{2}\,,
\end{eqnarray}

\noindent where $k_{\rm{A}}$ and $k_{\gamma}$ are the momenta of the outgoing axion and photon respectively, and $\alpha$ is the fine structure constant. The quantities $\mu_{0}$=0.88 and $\mu_{3}$=4.71 are the isoscalar and isovector nuclear magnetic moments respectively, given in nuclear magnetons. The parameter $\delta$ denotes the E2/M1 mixing ratio for this particular nuclear transition, while $\beta$ and $\eta$ are nuclear structure dependent ratios. Their values for the 14.4~keV de-excitation process of an $^{57}$Fe nucleus are $\delta$=0.002, $\beta=-1.19$, and $\eta=0.8$~\cite{haxtonlee}. Using these values in Eq.~(\ref{eq:axionphotonration}) we find

\begin{eqnarray} \label{eq:axionphotonratio2}
\frac{\Gamma_{\rm{A}}}{\Gamma_{\gamma}} = \left(\frac{k_{\rm{A}}}{k_{\gamma}}\right)^{3}1.82\;(-1.19g_{\rm{AN}}^{0}+g_{\rm AN}^{3})^{2}\,.
\end{eqnarray}

\noindent Introducing the effective nuclear coupling adapted to the case of $^{57}$Fe,  $ g_{AN}^{\rm eff} \equiv (-1.19g_{\rm{AN}}^{0}+g_{\rm AN}^{3})$, the corresponding axion flux at the Earth, as quoted in~\cite{bib:CAST_14keV}, is given by:

\begin{equation} 
\label{eq:axion_flux_14keV}
\Phi_{14.4}= \left(\frac{k_{\rm{A}}}{k_{\gamma}}\right)^{3}\times 4.56 \times 10^{23}\; (g_{AN}^{\rm eff})^{2} \; {\rm cm}^{-2}\,{\rm s}^{-1} .
\end{equation}

To take into account also non-relativistic limit, the factor $\left(\frac{k_{\rm{A}}}{k_{\gamma}}\right)^{3}$ has not been set equal to one. 
Using the expressions given in Eq.~(\ref{eq:ax_nucl_K}) and Eq.~(\ref{eq:ax_nucl_D}), it is possible to evaluate $g_{\rm AN}^{0}$ and $g_{\rm AN}^3$ for the DFSZ and KSVZ models and thus evaluate the corresponding fluxes for these two cases. Results show that the two fluxes are of the same order of magnitude (see~\cite{bib:cuore_RD} for a detailed discussion).

\subsubsection{Compton, bremsstrahlung and axio-RD processes}
\label{sec:compton}

The last solar production mechanisms explored in this paper arise if axions couple to electrons. The corresponding effective Lagrangian may be chosen as:

\begin{equation}
{\cal L}= i g_{Ae}\bar{\psi_e}\gamma_5\psi_e \phi_A
\label{eq:lagr_electron}
\end{equation}

\noindent where $g_{Ae}$ is the dimensionless axion-electron coupling constant. Axions can then be emitted within the Sun by the Compton process ($\gamma + e^-  \rightarrow e^- + A$) and by bremsstrahlung ($e^- + X  \rightarrow e^- + X + A$, where $X$ is an electron, a hydrogen or helium nucleus) occurring in the hot plasma. We also consider emission processes associated with the electron capture by an ion (axio-recombination), and to the bound-bound "axio-deexcitation": from the reevaluation by \cite{bib:cast_gae}, these processes lead to a non-negligible flux, which we call axio-RD. Since the derived fluxes scale in the same way as $g_{Ae}^2$, we take into account all these processes at the same time. For the axio-RD process we use a tabulated spectrum from~\cite{bib:cast_gae} (Fig.~\ref{fluxes}), while for the Compton-bremsstrahlung process we use the estimation~\cite{bib:cast_gae}:

\begin{equation}
\label{eq:CB_flux}
\begin{split}
\frac{d\Phi}{d{\text E}}& =  \left( \frac{d\Phi}{d{\text E}}\right)^{\rm Compton} + \left(\frac{d\Phi}{d{\text E}}\right)^{\rm bremsstrahlung}  \\
                                & = g_{Ae}^2\times 1.33\times 10^{33}\hspace{1mm}{\text E}^{2.987}\hspace{1mm}e^{-0.776\hspace{1mm}{\text E}} \\
                                & \quad + g_{Ae}^2 \times2.63\times 10^{35}\hspace{1mm}{\text E}\hspace{1mm} e^{-0.77\hspace{1mm}{\text E}}\frac{1}{1+0.667\hspace{1mm}{\text E}^{1.278}}
\end{split}
\end{equation}

\noindent where fluxes are in cm$^{-2}$ s$^{-1}$ keV$^{-1}$ and energies in keV.

While EDELWEISS data will be used to set model-independent constraints on $g_{Ae}$, valid for any ALP, explicit expressions for the coupling constant may be given for specific axion models.  Within the DFSZ axion models, where the coupling is at the tree level, we have:

\begin{equation}
\label{eq:beta_dfsz}
(g_{Ae})_{\rm DFSZ}=\frac{m_e}{3f_A} \cos ^2 \beta _{\rm DFSZ}
\end{equation}

\noindent where m$_e$ is the electron mass while $\tan \beta _{\rm DFSZ}$ was already defined in Section~\ref{sec:nucl_trans}. Here again, for model-dependent studies we will fix $\cos \beta_{\rm DFSZ}=1$. Therefore, in that case $g_{Ae}$ is numerically given by the expression:
\begin{equation} \label{eq:gaed}
(g_{Ae})_{\rm DFSZ} \simeq 1.68 \times 10^{-4} \frac{\rm GeV}{f_A} \simeq 2.84 \times 10^{-8} \frac{m_A}{\text{keV}}.
\end{equation}

In the KSVZ axion model characterized by the absence of tree-level coupling to electrons, $g_{Ae}$ is determined only by radiative corrections \cite{bib:srednicki}. As a consequence it is smaller than in the DFSZ model by a factor of about $\alpha^{2}$. The expression for this parameter is:

\begin{equation}
(g_{Ae})_{\rm KSVZ}=\frac{3\alpha^{2}Nm_e}{2\pi
f_{A}}\left(\frac{E}{N}\ln\frac{f_{A}}{m_e}-\frac{2}{3}\frac{4+z+w}{1+z+w}\ln\frac{\Lambda}{m_e}\right),\label{Gaee}
\end{equation}

\noindent where $E/N=0$ as discussed in Section~\ref{sec:primakoff_prod} for hadronic axions, and $\Lambda \sim 1$~GeV is associated with the QCD confinement scale. We therefore obtain numerically:

\begin{equation} \label{eq:gaex}
(g_{Ae})_{\rm KSVZ} \simeq -5.7 \times 10^{-7}\, \frac {\rm GeV}{f_A} 
\end{equation}

\subsection{Axions as dark matter}
\label{sec:dm_search}
Within the cosmological concordance $\Lambda$CDM model, a large fraction of the mass content of the universe is composed of dark matter (DM), the nature of which is still unknown. In particular the dynamics of our galaxy can be explained by the presence of a non-relativistic dark matter halo with a solar neighborhood density $\rho_{\rm DM}=0.3$~GeV/cm$^3$ in the conventional model. Axion-like particles are a possible candidate for dark matter, and the hypothesis of keV-scale ALPs was in particular proposed as an explanation of the annual modulation observed by DAMA in NaI crystals~\cite{bib:DAMA_dmaxion}.

When testing this scenario, we will assume that axions constitute all of the galactic dark matter. The total, average flux of dark matter axions on Earth is then:

\begin{equation}
\Phi_{\rm DM} [/ \rm cm^{2}/\rm s]= \rho_{\rm DM} \cdot v_A / m_A = 9.0\times 10^{15} \,\left(\frac{\rm keV}{m_A}\right) \cdot \beta
\label{eq:DM_flux}
\end{equation}

\noindent In this expression, m$_A$ is the axion mass and $v_A$  the mean axion velocity distribution with respect to the Earth, $\beta \simeq 10^{-3}$. The flux does not depend on any axion coupling.



\section{Axion interactions in EDELWEISS detectors}
\label{sec:detection}

\begin{figure}[!ht]
\centering
\includegraphics[width=\textwidth]{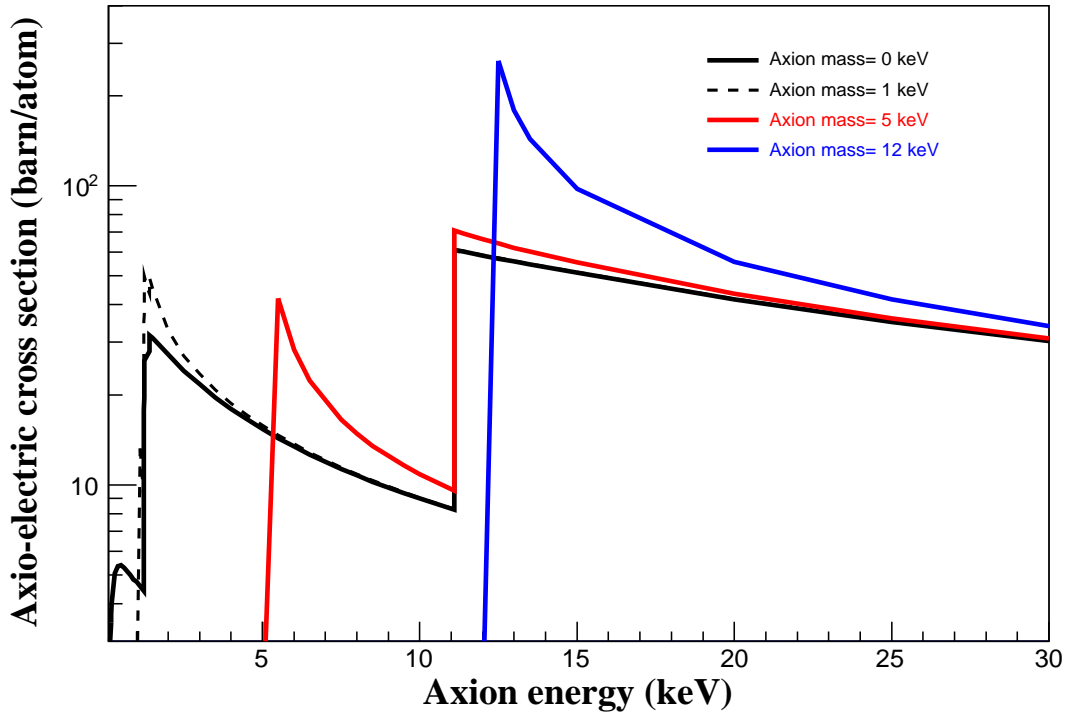}
\caption{Axio-electric cross section for different axion masses, computed for germanium and normalized with $g_{Ae}=1$. The discontinuities at $1.2-1.4$~keV and 11.1~keV are due to electron shell energies.}
\label{axio_cross}
\end{figure}

In this paper we will use, depending on the production channel,  two different mechanisms for axion detection.
\begin{itemize}
\item Through the Primakoff effect, axions can be converted into photons in the intense electric field of the germanium crystal~\cite{bib:primakoff_theory}. The wavelength of relativistic solar axions, with an energy of a few keV, is of the same order of magnitude as the inter-atomic spacing of the detector. Therefore, depending on the direction of the incoming axion flux with respect to the lattice, the axion signal can be enhanced significantly through Bragg diffraction (EDELWEISS detectors are mono-crystals). The corresponding correlation of the count rate with the position of the Sun in the sky also helps further with an effective background rejection. We can express the Bragg condition as a function of the axion energy, neglecting the axion mass and the target recoil: $E_A=\vert{\text{\bf G}^2}\vert/(2{\bf u}\cdot{\bf G})$, where ${\bf G}$ is a reciprocal lattice vector and {\bf u} is a unit vector directed towards the Sun. For non-zero axion masses, the Bragg condition is changed by the axion dispersion relation and becomes, for $m_A \ll E_A$:

\begin{equation}
E_A^2 = \frac{{\bf G}^4}{4({\bf u}\cdot {\bf G})^2} + \frac{m_A^2}{2}
\label{eq:bragg_condition}
\end{equation}

\noindent For $m_A=200$~eV, the relative correction on $E_A$ with respect to the case of massless axions is $m_A^2/4E_A^2 \sim 10^{-2}\,E_A$.

\item Axions can also be detected through the axio-electric effect, the equivalent of a photo-electric effect with the absorption of an axion instead of a photon: A+e$^-$+Z$\rightarrow$ e$^-$+Z. The axio-electric cross-section as a function of the axion energy was computed in~\cite{Derevianko, Pospelov,bib:cuore_RD}, and is represented for several values of its mass in Fig.~\ref{axio_cross}:


\begin{equation}
\label{eq:ae_cross_section}
\sigma_{Ae}(E)=\sigma_{\text{pe}} (E) \frac{g_{Ae}^2}{\beta}\frac{3E^2}{16\pi\alpha m_e^2}\left(1-\frac{\beta^{\frac{2}{3}}}{3}\right)
\end{equation}

In this expression $\sigma_{\text{pe}}$ is the germanium photoelectric cross-section, taken from~\cite{bib:photoel} , $\beta$ is the ratio of the axion velocity to the speed of light, $\alpha$ is the fine structure constant and $m_e$ the electron mass. Through the axio-electric effect, an incoming axion of energy $E$ (relativistic or not) will generate an electron recoil with the same energy within an EDELWEISS detector.

\end{itemize}

\section{EDELWEISS-II data and backgrounds}
\label{sec:experiment}

EDELWEISS was primarily designed as a direct detection WIMP (Weakly Interacting Massive Particle) search experiment, in which nuclear recoils induced by WIMPs from the galactic halo are detected using germanium detectors working at very low temperatures (18~mK). We simultaneously measure ionization and phonon signals. The comparison of the two signals allows a separation on an event-by-event basis of nuclear recoils from electron recoils induced by $\beta$ and $\gamma$ radioactivity. These electron recoils constitute the major source of background in most present-day direct WIMP searches. In particular, the ten 400-g EDELWEISS-II detectors used in this analysis, called InterDigit (ID), are equipped with a set of thin aluminium interleaved electrodes. This makes possible the identification of near-surface interactions based on the measured ionization signals~\cite{id_paper}, and therefore the definition of an inner, so-called fiducial volume for each detector. Detectors are operated at the Underground Laboratory of Modane (LSM) in the Frejus Tunnel under the French-Italian Alps. They are protected from external radioactivity by lead and polyethylene shields and also an active muon veto. EDELWEISS-II has provided frontier sensitivities to WIMP-nucleon cross-sections for WIMP masses above 50 GeV~\cite{ref_EDW2first,ref_EDW2}, as well as for low-mass WIMPs $\sim 10$~GeV~\cite{lowmasspaper}.

Through their weak coupling either to photons or to electrons, axions generate electron recoils which can be detected in the EDELWEISS bolometers. The axion searches presented here are based on data collected with ten ID bolometers during 14 months in 2009-2010, as described in~\cite{ref_EDW2first, ref_EDW2}. The offline pulse reconstruction and calibration are identical to~\cite{ref_EDW2first, ref_EDW2}.
Axion-induced events were searched for within the population of low-energy fiducial, electronic recoils:
\begin{itemize}
\item Fiducial events were selected by requiring the absence of any signal above 4 sigma on the veto and
 guard electrodes and by constraining the difference in the measured values of the two collecting electrodes. The efficiency of this cut results in a fiducial mass of 160~g for each detector~\cite{ref_EDW2first}.
\item For each fiducial event, we measured both the heat energy $E_{\rm heat}$ and a fiducial ionization energy $E_{\rm ion}$, based on the combination of signals from both collecting electrodes. Fiducial electron recoils are gaussian distributed along the line $E_{\rm ion}=E_{\rm heat}$. We rejected events beyond three standard deviations from this line.
\end{itemize}

An exposure adapted to this event selection was defined: we discarded time periods with noisy fiducial ionization or heat signals, and obtained a homogeneous data set for each detector with 280 live days on average. The total exposure reaches 448~kg.d for this analysis. It is larger than the WIMP-search exposure published in~\cite{ref_EDW2} 
mainly because the axion search does not depend as strongly on the purity of the fiducial selection, and requirements on the resolution on the heat and ionization guard signals can be relaxed. To reject misreconstructed pulses, a cut was applied on the $\chi^2$ of the fit to heat pulse shapes, keeping 98.7~\% efficiency. We also rejected coincidence events in neighboring bolometers and events detected in the muon veto. The latter has negligible deadtime~\cite{bib:edw_muons}.

At low energies, an efficiency loss $\epsilon_{\text online}$ appears because of the online trigger. The efficiency function was computed from our knowledge of the time variations of this trigger. It was cross-checked with gamma calibrations. In addition, we selected events with both heat and fiducial ionization above a given threshold defined on a per-detector basis.
The cut on fiducial ionization is essential to remove the large number of heat-only pulses recorded during the experiment. They are due to lead recoils associated with surface radioactivity, internal radioactivity of the heat sensors and mechanical noise.

Finally, for each selected event we combined the heat and fiducial ionization to obtain an optimal energy estimator for fiducial electron recoils: $\tilde{E} = w_{\rm heat}\,E_{\rm heat}+w_{\rm ion}\,E_{\rm ion}$ where $w_{\rm heat}$ and $w_{\rm ion}$ are weights depending on the heat and fiducial ionization resolutions. An analysis threshold on $\tilde{E}$ is set for each detector, by requiring that $\epsilon_{\text online}(\tilde{E})>50\%$ and $\epsilon_{\text other\hspace{1 mm}cuts}(\tilde{E})>95\%$. With these cuts, three detectors have a threshold at 2.5~keV, two at 3~keV and five at 3.5~keV. The average FWHM at low energy is 0.8~keV for $\tilde{E}$.

\begin{figure}[!ht]
\centering
\includegraphics[width=0.48\textwidth]{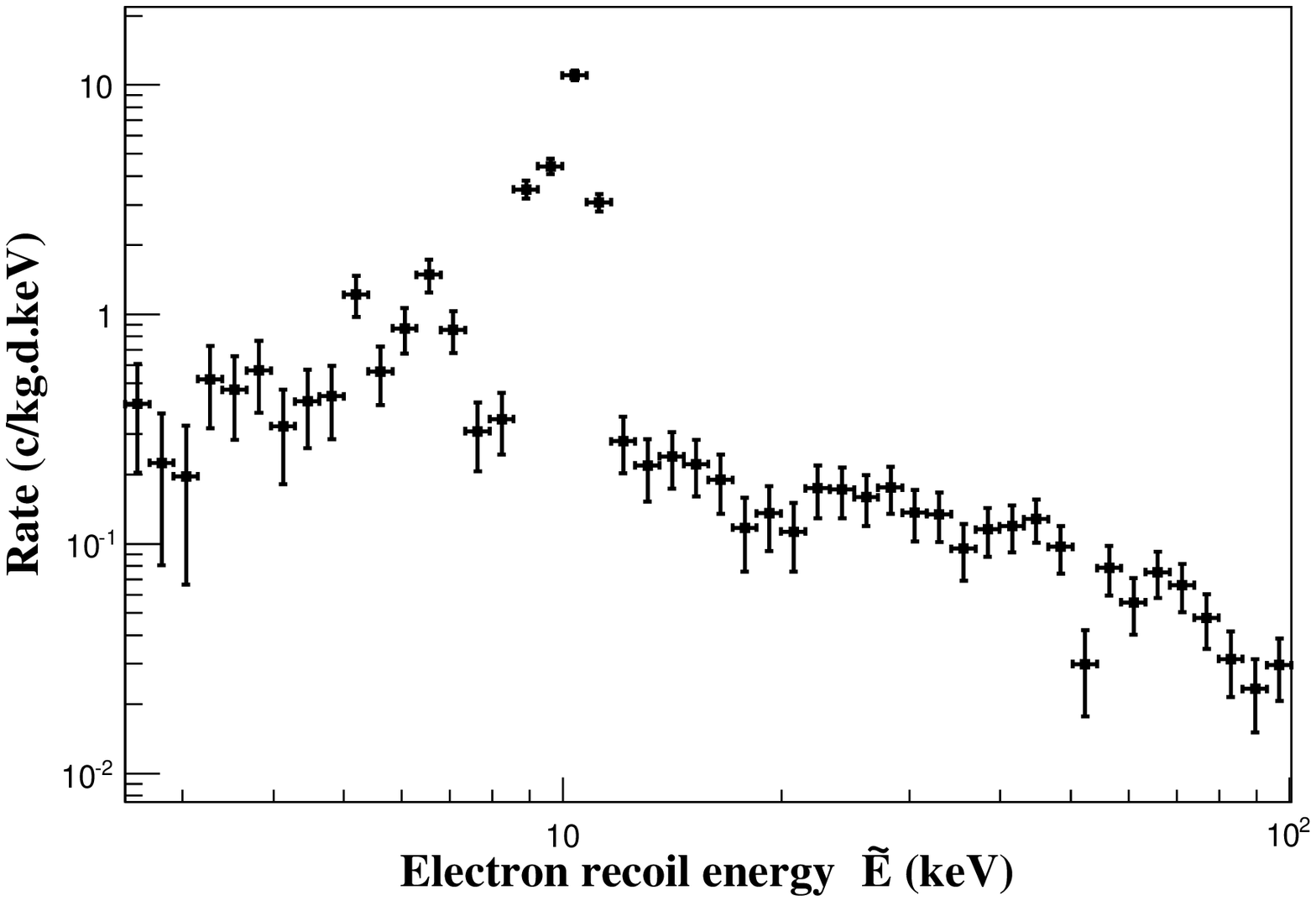}
\includegraphics[width=0.48\textwidth]{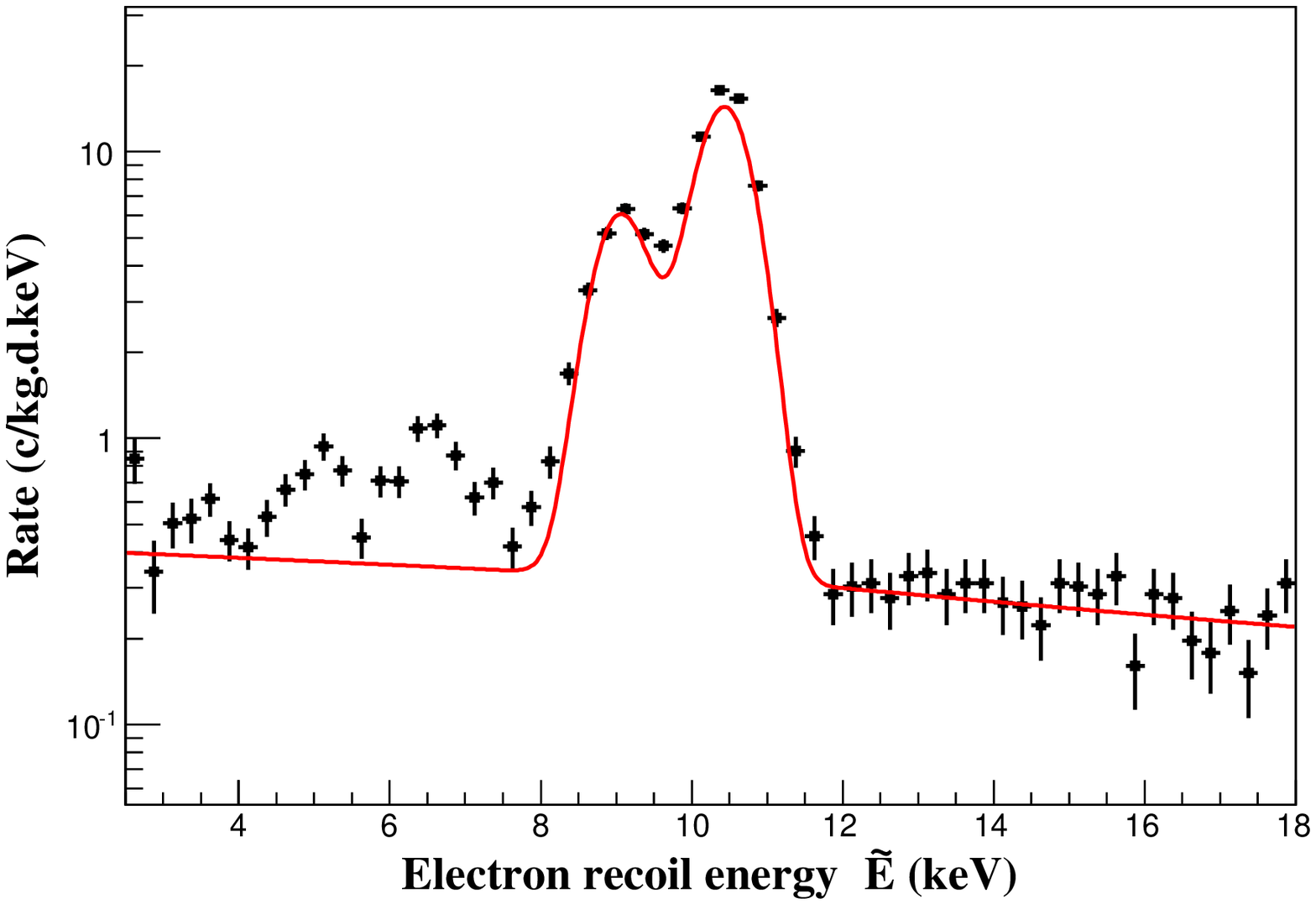}
\caption{Left: Efficiency-corrected electron recoil spectrum in the fiducial volume of a single bolometer called ID3, in the energy range $2.5-100$~keV. The smooth Compton feature is visible as well as low-energy lines from induced radioactivity and cosmogenic activation of germanium. Right: Stacked, efficiency-corrected electron recoil spectrum for the full exposure in the $2.5-18$~keV range. The red line is the background model $B(\tilde{E})$ used in all analyses but Primakoff: a smooth Compton component linearly extrapolated below 12~keV, together with 10.37~keV and 8.98~keV cosmogenic lines.}
\label{spectrum_data}
\end{figure}

Fig.~\ref{spectrum_data} (left) shows the event rate as a function of $\tilde{E}$ for a typical detector, after the data selection described above and after correction by the online trigger efficiency function. The background consists of a Compton profile with a smooth, slightly decreasing energy dependence, together with radioactive peaks, notably identified at 10.37~keV ($^{68}$Ge), 9.66~keV ($^{68}$Ga), 8.98~keV ($^{65}$Zn), 6.54~keV ($^{55}$Fe), 5.99~keV ($^{54}$Mn) and 4.97~keV ($^{49}$V). An additional smooth background increase at low energy may also be expected from unrejected surface interactions.
 In the axion searches described below, two different background models are used depending on the search:

\begin{enumerate}
\item \textit{Primakoff solar axions.}
In this study, we will exploit the time and energy dependence of the axion signal to quantify $g_{A\gamma}$. This results in an effective background rejection of about two orders of magnitude~\cite{Cebrian}. Furthermore the expected global energy distribution of the signal has a larger width than the detector FWHM. As a consequence we will include all radioactive peaks in the background model used for this analysis, in addition to a smooth component. The smooth time variation of these peaks is negligible with respect to the sharp and fast-varying axion signal. This analysis also requires that we first study each detector individually: a background model is adjusted to each detector spectrum.

\item \textit{Other axion searches.}
In all other searches, the axion signal has no time dependence to first order and is simply  identified by its spectral shape (such as a line) in the stacked spectrum of all detectors. The time evolution of the 8.98 keV and 10.37 keV line intensities (with decay times of 243 and 271 days, respectively) allowed us to measure the intensity of these specific cosmogenic lines and include them in the background model, independently from a potential time-independent axion signal at the same energy. However, the other radioactive peaks cannot be confidently estimated from their decrease with time for lack of statistics. Therefore they are conservatively not included in the background model. Above 12~keV, the smooth component of the spectrum is adjusted by a polynomial fit. In this energy region, an increase of the count rate is observed when energy decreases as expected from simulations~\cite{bib:EDW_bkg}. The smooth component is extrapolated below 12~keV with a linear fit. Two detectors, having a significantly higher background than the others below 8~keV~\cite{lowmasspaper}, were discarded from the stacked spectrum below this energy. This results in an effective exposure of 357~kg.d at low energy for these analyses. Fig.~\ref{spectrum_data} (right) shows the stacked, efficiency-corrected background rate of the detectors used in the analysis, together with the associated background model. We will refer to this model as $B(\tilde{E})$ in the following.
 

\end{enumerate}

\section{Axion search: Primakoff solar axions}
\label{sec:primakoff}

We consider the scenario in which axions are produced in the Sun by inverse Primakoff conversion, resulting in the flux given in Eq.~(\ref{eq:solar_prim_flux}), and are detected again by the Primakoff effect, using coherent Bragg diffraction. This axion search relies only on the existence of an effective axion-photon coupling. Using the same formalism as in~\cite{Cebrian}, the expected count rate in a single detector as a function of energy, time, and the detector orientation $\alpha$ is given by:

\begin{equation}
\begin{split}
R(\tilde{E},t,\alpha)&=2(2\pi)^3\,\frac{V}{v_a^2}\sum_G\frac{d\phi}{dE_A}\,\frac{g_{A\gamma}^2}{16\pi^2}\,\sin(2\theta)^2\frac{1}{|{\bf G}|^2}\left|S({\bf G})F_A^0({\bf G})\right|^2\hspace{2pt}W(E_A,\tilde{E})\\
&=\left(\frac{g_{A\gamma}\times10^8}{{\text{GeV}}^{-1}}\right)^4\overline{R}(\tilde{E},t,\alpha) \;\; \equiv \;\;\lambda \overline{R}(\tilde{E},t,\alpha) .
\end{split}
\end{equation}

\begin{figure}[!ht]
\centering
\includegraphics[width=\textwidth]{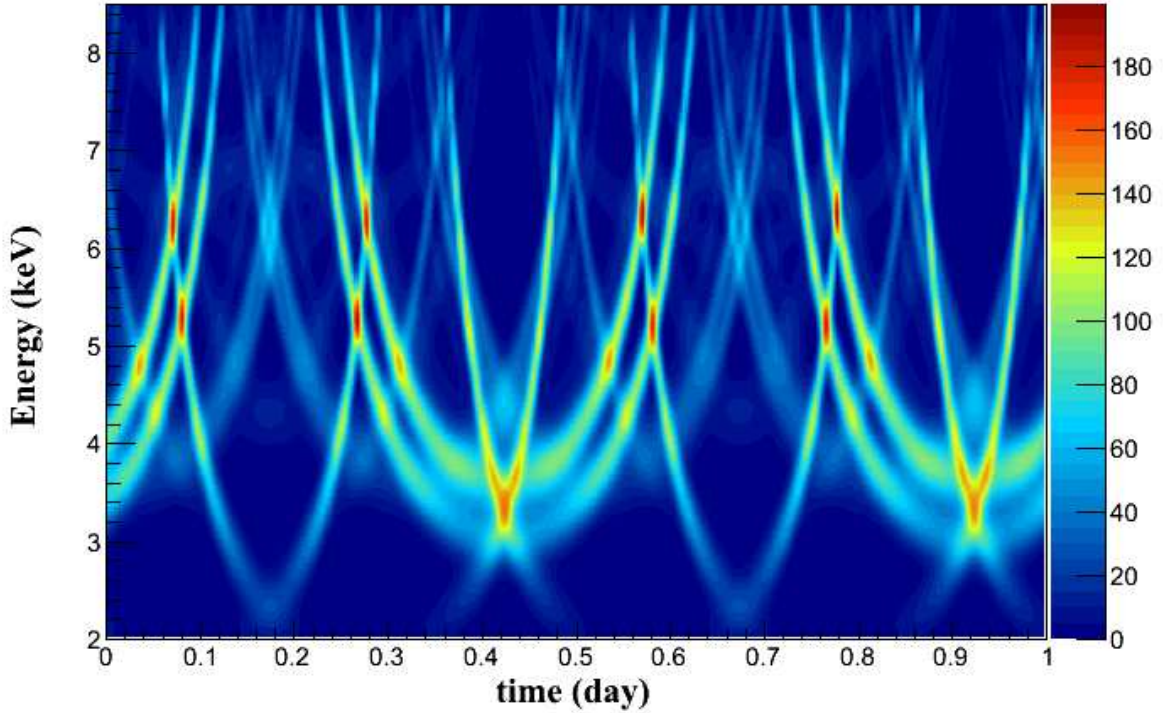} 
\caption{Example of the theoretical Primakoff conversion signal in a single detector, in counts per kg.d.keV, for a detector resolution of 0.5~keV and $g_{A\gamma}=10^{-8}$ GeV$^{-1}$.}
\label{2Drate}
\end{figure}

\noindent In the first expression, $V$ is the detector volume and $v_a$ is the volume of the elementary cell of the crystal lattice. The sum is over the vectors $\bf G$ of the reciprocal lattice, $2\theta$ is the scattering angle (related to the time-varying direction of the Sun with respect to $\bf G$), S is the structure factor and $F_A^0$ is the atomic form factor associated with the electrostatic field. The function $W$ represents the detector resolution for the observable energy $\tilde{E}$, $E_A$ is the axion energy related to ${\bf G}$ by the Bragg condition. Fig.~\ref{2Drate} illustrates both time and energy variation of the signal, for a given detector orientation in local (terrestrial) coordinates. Note that this expression for the rate still applies for mildly relativistic axions, after the appropriate modifications of both the Bragg condition giving $E_A$ and the solar flux are taken into account. Up to $m_A\simeq 200$~eV, the solar Primakoff flux changes by less than 1~\% and the Bragg energy is shifted by $\sim 1$~\%, a negligible value with respect to detector resolution hence the expression remains valid.

The geodesic location of the Underground Laboratory of Modane is (45.14$^\circ$ N, 6.68$^\circ$ E). While the vertical axis of the bolometer tower is aligned with the [001] axis of each detector, with a precision of about one degree, the individual azimuthal orientation $\alpha$ of each detector was not measured. Therefore, in a first step we will adapt the method developed in the same context by~\cite{Cebrian} for a single bolometer. We use the following time correlation function assuming a given orientation $\alpha$:

\begin{equation}
\chi_k(\alpha) = \epsilon_k\sum_i\left[ \overline{R_k}(t_i)-\left\langle\overline{R_k}\right\rangle\right]\cdot n_{ik}\equiv  \sum_i^nW_{ik}\cdot n_{ik}
\end{equation}

~\\where $\epsilon_k$ is the detector efficiency, $n_i$ indicates the number of measured events in the time interval $[t_i, t_i+\Delta t]$, the index $k$ refers to the energy interval $[\tilde{E}_k, \tilde{E}_k+\Delta \tilde{E}]$
and the sum is over the total period of data taking. We use the analysis window $3-8$~ keV,
which contains most of the expected signal. The Dirac-like brackets indicate an average over time. The distribution of $n_i$ is given by a Poisson distribution with average:

\begin{equation}
\left\langle n_{ik} \right\rangle= \epsilon_k\left[\lambda\overline{R_k}(t_i)+b_k\right]\Delta t\Delta \tilde{E}
\end{equation}

\noindent where $b_k$ is the individual, constant detector background in the considered energy interval. Assuming background dominates, we compute\footnote{This corrects the expression found in~\cite{Cebrian}.}:

\begin{equation}
\begin{split}
        &\left\langle \chi_k \right\rangle =\epsilon_k^2\,\lambda\sum_iW_{ik}^2\,\Delta t\,\Delta \tilde{E}\equiv \lambda\cdot A_k\\
        &\sigma^2(\chi_k)\approx \epsilon_k \,b_k\,A_k
\end{split}
\end{equation}

\noindent Minimizing the associated likelihood function, we derive a simple estimator for the reduced coupling $\lambda$:

\begin{equation}
        \tilde{\lambda}(\alpha) = \frac{\sum_k\frac{\chi_k}{\epsilon_kb_k}}{\sum_k\frac{A_k}{\epsilon_k b_k}}    
\end{equation}



~\\In order to combine all the detectors and at the same time take into account the lack of knowledge of the  azimuthal orientation of each crystal, we apply the following procedure. Combining all detectors and scanning over all possible orientations, we obtain from the data an overall distribution for $\tilde{\lambda}$, $D_{\text{real data}}(\tilde{\lambda})$. We performed Monte Carlo simulations including the detector exposures, efficiencies and backgrounds as well as a potential axion signal. These simulations show that, in the presence of an axion signal, this distribution D charts a tail at high $\tilde{\lambda}$. 
Based on simulations, we therefore introduce a statistical observable $I$ given by:

$$I = \int_{|\tilde{\lambda}| < \tilde{\lambda}_c} D(\tilde{\lambda}) - \int_{|\tilde{\lambda}| > \tilde{\lambda}_c} D(\tilde{\lambda})$$

\noindent where $\tilde{\lambda}_{\text{cutoff}}=0.003$.
The simulations allow us to obtain the expected distribution of $I$  for a given $\lambda_0$ and set of detector orientations\footnote{Note that this procedure is independent of the chosen orientation of the detectors.} $\alpha_0^{\text{bolo}}$. 
The measured value $I_{\text{real data}}$ is compatible with simulations carried out for $\lambda_0=0$. By scanning over $\lambda_0$ and comparing the resulting distributions of $I$ with $I_{\text{real data}}$, we can place a 95\% CL upper limit on the axion-photon coupling:

$$ g_{A\gamma} < 2.13\times10^{-9}\, {\rm GeV}^{-1}  \;\;\;{\rm (95\% CL)}.$$

\begin{figure}[!ht]
\centering
\includegraphics[width=\textwidth]{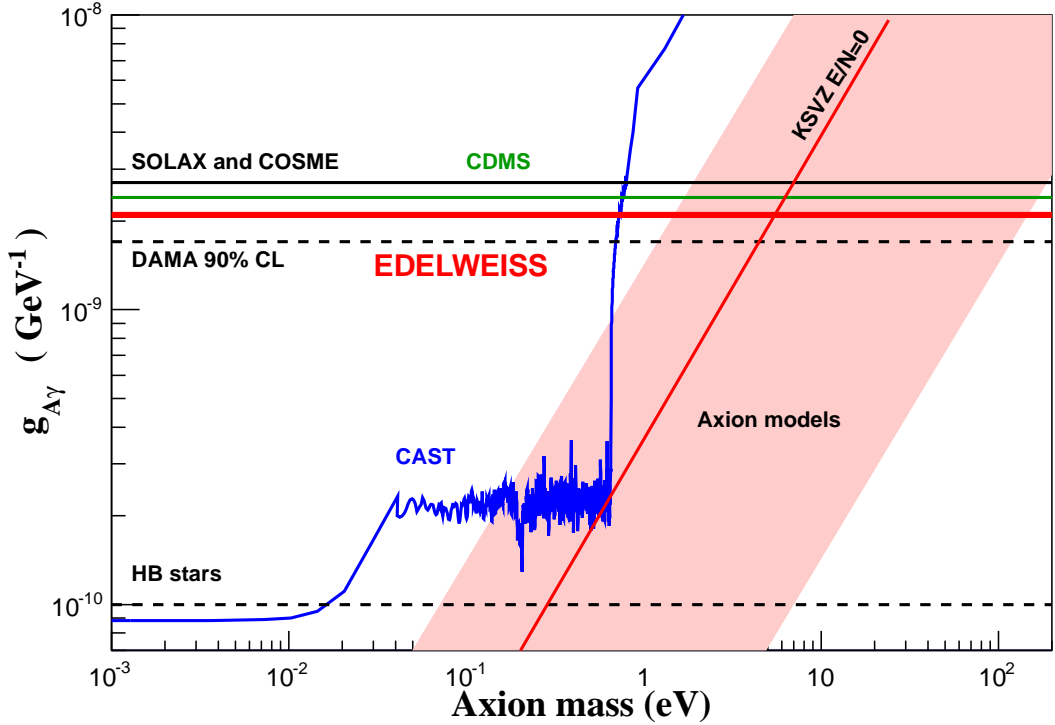}
\caption{95~\% CL limit on the $g_{A\gamma}$ coupling from the solar Primakoff flux obtained by EDELWEISS-II (red), compared to other crystal experiments such as SOLAX~\cite{bib:SOLAX}, COSME~\cite{bib:COSME}, CDMS~\cite{bib:CDMS} (green) and DAMA~\cite{bib:DAMA_gay} (90~\% CL)). We also show the CAST limit~\cite{bib:CAST_gay} (blue) and indirect bounds from HB stars~\cite{bib:hbstars}. The light red band labeled `Axion models' represents typical theoretical models with $\mid E/N-1.95 \mid$ =0.07-7. The red solid line inside this band represents the case E/N=0 (KSVZ model).}
\label{limit_gay}
\end{figure}

\noindent This limit is shown in Fig.~\ref{limit_gay} and compared with constraints from other experiments and astrophysical bounds. These results will be discussed
in more details in Section~\ref{results}.

\section{Axion search: 14.4 keV solar axions}

\begin{figure}[!ht]
\centering
\includegraphics[width=\textwidth]{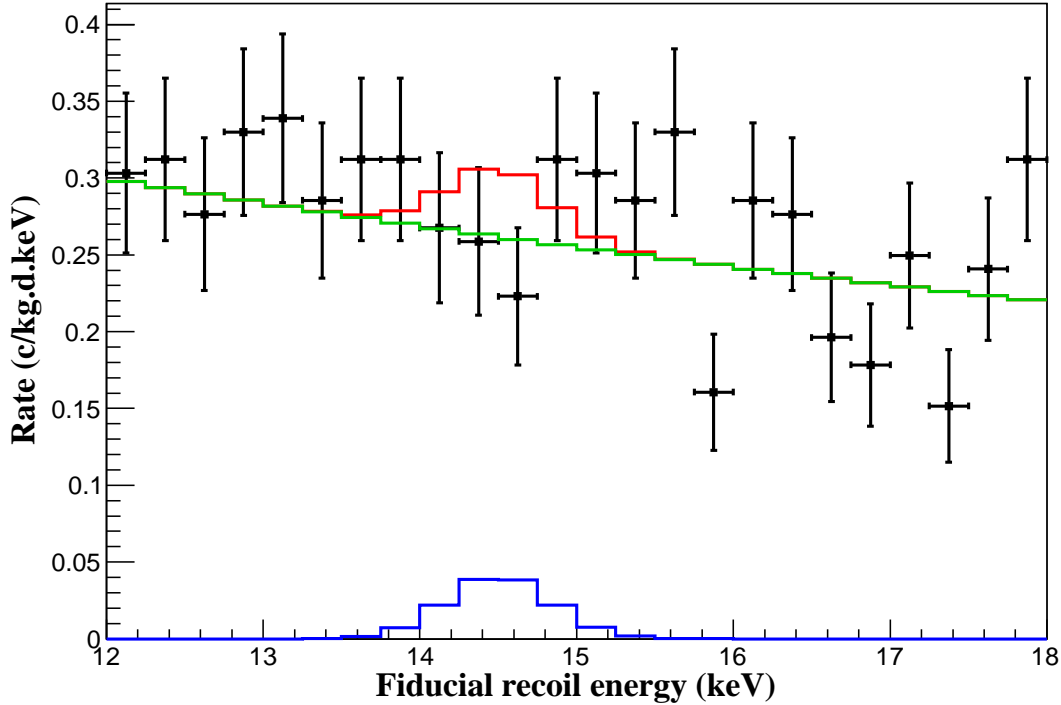}
\caption{Stacked electron recoil spectrum around 14.4~keV. Blue: detector response to 14.4~keV solar axions using axio-electric conversion at the 90\% CL limit. Green: background model. Red: axion signal with coupling at the 90\%~CL limit superimposed on the background model.}
\label{stack_14keV}
\end{figure}

We now test the scenario in which solar axions are produced in the $^{57}$Fe magnetic transition and detected by the axio-electric effect in a Ge crystal, resulting in a 14.4~keV electron recoil. The expected rate in counts per keV is the product of the flux $\Phi_{14.4}$ (Eq.~\ref{eq:axion_flux_14keV}), $\beta=v/c$, the axio-electric cross section given in Eq.~(\ref{eq:ae_cross_section}), the individual detector resolution $\sigma_i$ and exposure $M_iT_i$, summing over all detectors $i$:
\begin{equation}
\label{eqn:rate_14kev}
\begin{split}
R_{\text{14.4}}(\tilde{E})&=\beta^3\,\Phi_{14.4}\,\sigma_{\text{A}}(14.4)\sum_iM_iT_i\frac{1}{\sqrt{2\pi}\sigma_i}\times e^{-\frac{(\tilde{E}-14.4)^2}{2\sigma_i^2}}\\
&\equiv\lambda\times\overline{R}_{\text{14.4}}(\tilde{E})\quad\quad{\text{where }} \lambda=(g_{Ae}\times g_{AN}^{\rm eff})^2\\
\end{split}
\end{equation}
\noindent At 14.4~keV, the online trigger efficiency is equal to 1 for all 10 detectors. Fig.~\ref{stack_14keV} shows the stacked electron recoil spectrum in the $12-18$~keV interval.
There is no hint of a line at 14.4 keV and we therefore derive a limit on the line intensity using a binned likelihood function and assuming Poisson statistics for the background:

\begin{equation}
\label{likelihood}
L=\prod_i e^{-N^{\text{th}}_i}\frac{(N^{\text{th}}_i)^{N^{\text{exp}}_i}}{N^{\text{exp}}_i!}
\end{equation}

\noindent Here  $N^{\text{exp}}_i$ is the observed number of events in the energy bin $i$ and  $N^{th}(\tilde{E})=\lambda\overline{R}_{\text{14.4}}(\tilde{E})+B(\tilde{E})$.
~\\ The likelihood function is gaussian to a very good approximation. In order to deal with possible negative background fluctuations, we use the following prescription. If the likelihood best fit is positive we use a standard gaussian 90\% upper limit,  the value is negative we assume it is equivalent to a zero measurement. This is a conservative approach which solves issues of undercoverage and empty intervals as discussed in~\cite{Cousins}. We find $R_{\text{14.4}}<0.038$ counts/kg/d. This method was validated with MC simulations.

For a low-mass axion, this result translates to a 90\% CL constraint on the couplings: 

$$g_{AN}^{\rm eff} \times g_{Ae}<4.7\times 10^{-17}
$$

\begin{figure}[htbp]
\begin{center}
{\includegraphics[width=\textwidth]{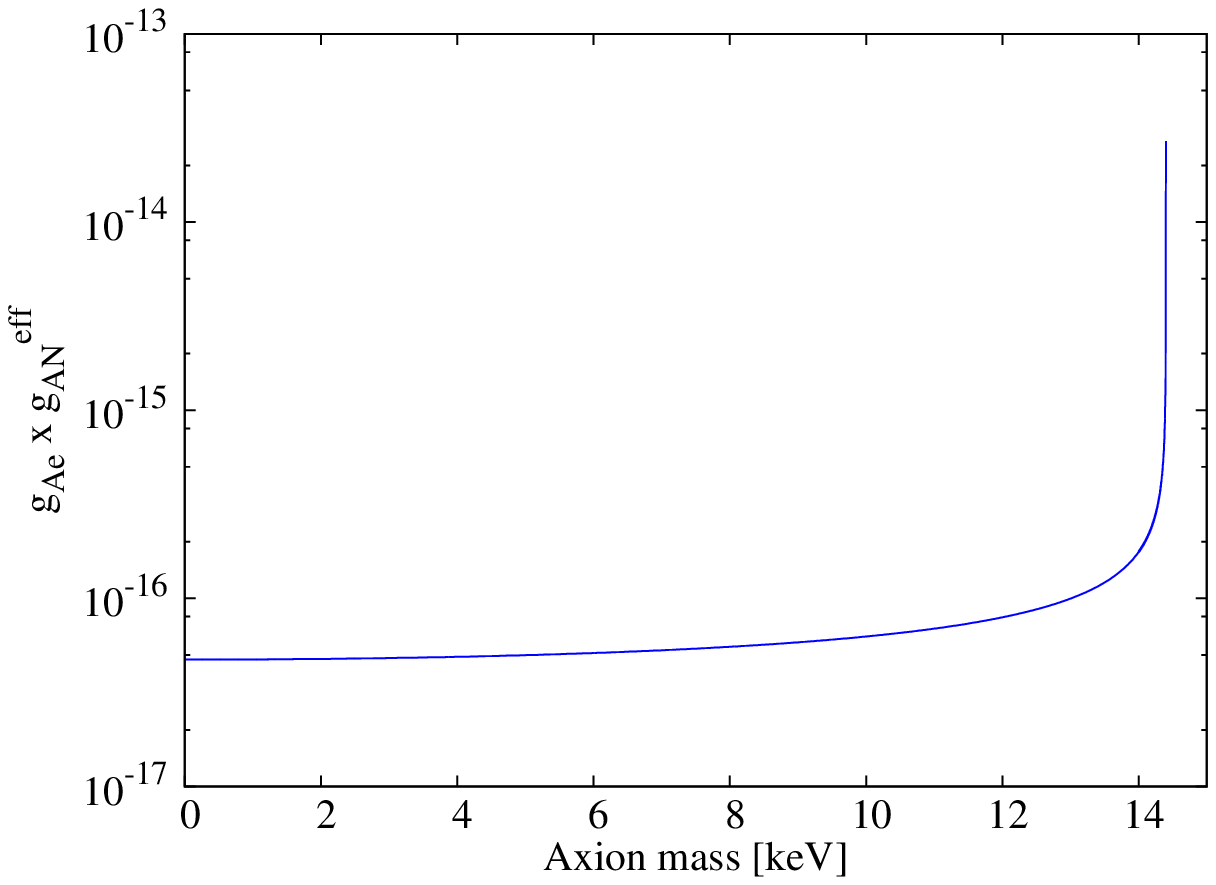}}
\end{center}
\caption{90~\% CL upper limits for $ g_{Ae} \times g_{AN}^{\rm eff} $ as a function of the axion mass $m_A$ obtained with the EDELWEISS-II data.}
\label{fig:gae-gan}
\end{figure}

\noindent Using the relationships given in Eq.(\ref{eq:axion_flux_14keV}) and Eq.(\ref{eq:ae_cross_section}), it is possible to obtain the upper limits for $g_{AN}^{\rm eff} \times g_{Ae}$ as a function of the axion mass $m_A$ for axion masses up to 14~keV. Fig.~\ref{fig:gae-gan} shows this model independent limit.

\section{Axion search: Compton, bremsstrahlung and axio-RD}

In this scenario, solar axions are produced through Compton, bremsstrahlung and axio-RD processes, resulting in the flux given by Eq.~(\ref{eq:CB_flux}). These axions can be detected by the axio-electric effect in the crystal, resulting in an expected count rate given by:

\begin{equation}
\begin{split}
R_{\text{C-B-RD}}(\tilde{E})&=\int dE_A\, \sigma_{\text{A}}(E_A)\left(\frac{d{\Phi}^{\text{C-B-RD}}}{dE_A}\right)
\times \sum_i\epsilon_i(\tilde{E})M_iT_i\frac{1}{\sqrt{2\pi}\sigma_i}\times e^{-\frac{(\tilde{E}-E_A)^2}{2\sigma_i^2}}\\
&\equiv \lambda\times\overline{R}_{\text{C-B-RD}}(\tilde{E})\quad\quad{\text{where }} \lambda= g_{Ae}^4\\
\end{split}
\end{equation}

\noindent The notations are identical to Eq.~(\ref{eqn:rate_14kev}), $\epsilon_i$ being the efficiency function for a given detector $i$, which is relevant at low energy. We look for Compton-bremsstrahlung-axio-RD solar axions in the $2.5-30$~keV energy window, which contains most of the signal (see Fig.~\ref{fluxes}).  We use the same likelihood procedure described in the previous section, where $N^{th}(\tilde{E})$ becomes:

\begin{equation}
N^{th}(\tilde{E})=\lambda\overline{R}_{\text{C-B-RD}}(\tilde{E})+B(\tilde{E})
\quad\quad(\lambda=g_{Ae}^4)
\end{equation}

\begin{figure}[!ht]
 \centering
\includegraphics[width=\textwidth]{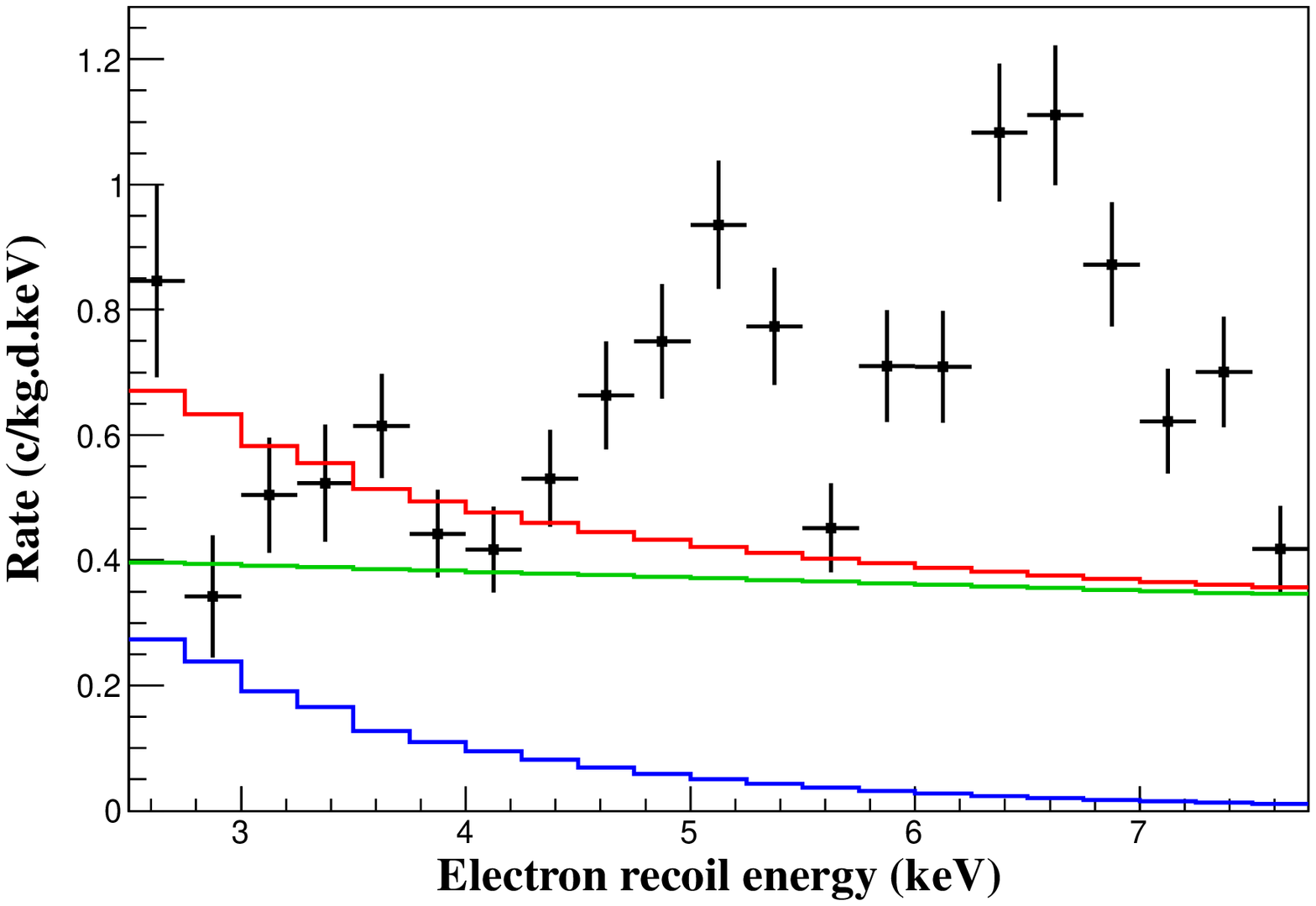}
\caption{Efficiency-corrected stacked electron recoil spectrum for the whole exposure, close to the analysis threshold. The detector response for a Compton Bremsstrahlung axio-RD, zero-mass axion signal at the 90\% confidence limit is represented by the blue curve, while the green curve shows the conservative background model. Red curve: signal superimposed over the background model. Note that the number of detectors used to compute the spectrum depends on the energy range considered.}
\label{stack_CB}
\end{figure}

The expected signal is not a line feature, so the likelihood analysis window has been tailored to each axion mass. Fig.~\ref{stack_CB} shows an example for $m_A=0$~keV, for which the expected signal stands mostly below 5~keV, and therefore the likelihood is not strongly affected by the presence of activation peaks. For $m_A>5$~keV, the presence of the peaks, not included in the background model $B(\tilde{E})$, degrades the sensitivity to axions. For each axion mass, a 90~\% CL limit is found using the same prescription as before. The constraint found for axions with $m_A \ll$~few keV is $R_{\rm C-B-RD}<0.46$ counts/kg/day. This translates to a constraint on the axio-electric coupling: $g_{Ae}<2.56\times 10^{-11}$ at 90\% CL. The evolution of the limit on $g_{Ae}$ as a function of $m_A$ can be found on Fig.~\ref{limit_gae}.
 
\section{Axion search: dark matter axions}
\label{sec:dm}

We now focus on the scenario in which axions constitute the entire dark matter halo of our galaxy. Since the galactic DM is non-relativistic, the resulting signal due to the axio-electric coupling will consist in electron recoils at an energy equal to the axion mass $m_A$. From Eq.~(\ref{eq:DM_flux}) and Eq.~(\ref{eq:ae_cross_section}), the expected axion count rate is:

\begin{equation}
\begin{split}
R_{\text{DM}}(\tilde{E})&=\Phi_{\text{DM}}\sigma_{\text{A}}(m_A)\times\sum_i\epsilon_i(\tilde{E})M_iT_i\frac{1}{\sqrt{2\pi}\sigma_i}\times e^{-\frac{(\tilde{E}-m_A)^2}{2\sigma_i^2}}\\
& = \lambda\times\overline{R}_{\text{DM}}(\tilde{E})\quad\quad{\text{where }} \lambda= g_{Ae}^2\\
\end{split}
\end{equation}

~\\The notations are the same as above. We look for galactic axions in the [2.5 keV - 100 keV] mass window. We proceed with a binned likelihood as in Eq.~(\ref{likelihood}), where

\begin{equation}
N^{th}(\tilde{E})=\lambda\overline{R}_{\text{DM}}(\tilde{E})+B(\tilde{E})
\end{equation}

\begin{figure}[!ht]
\centering
\includegraphics[width=\textwidth]{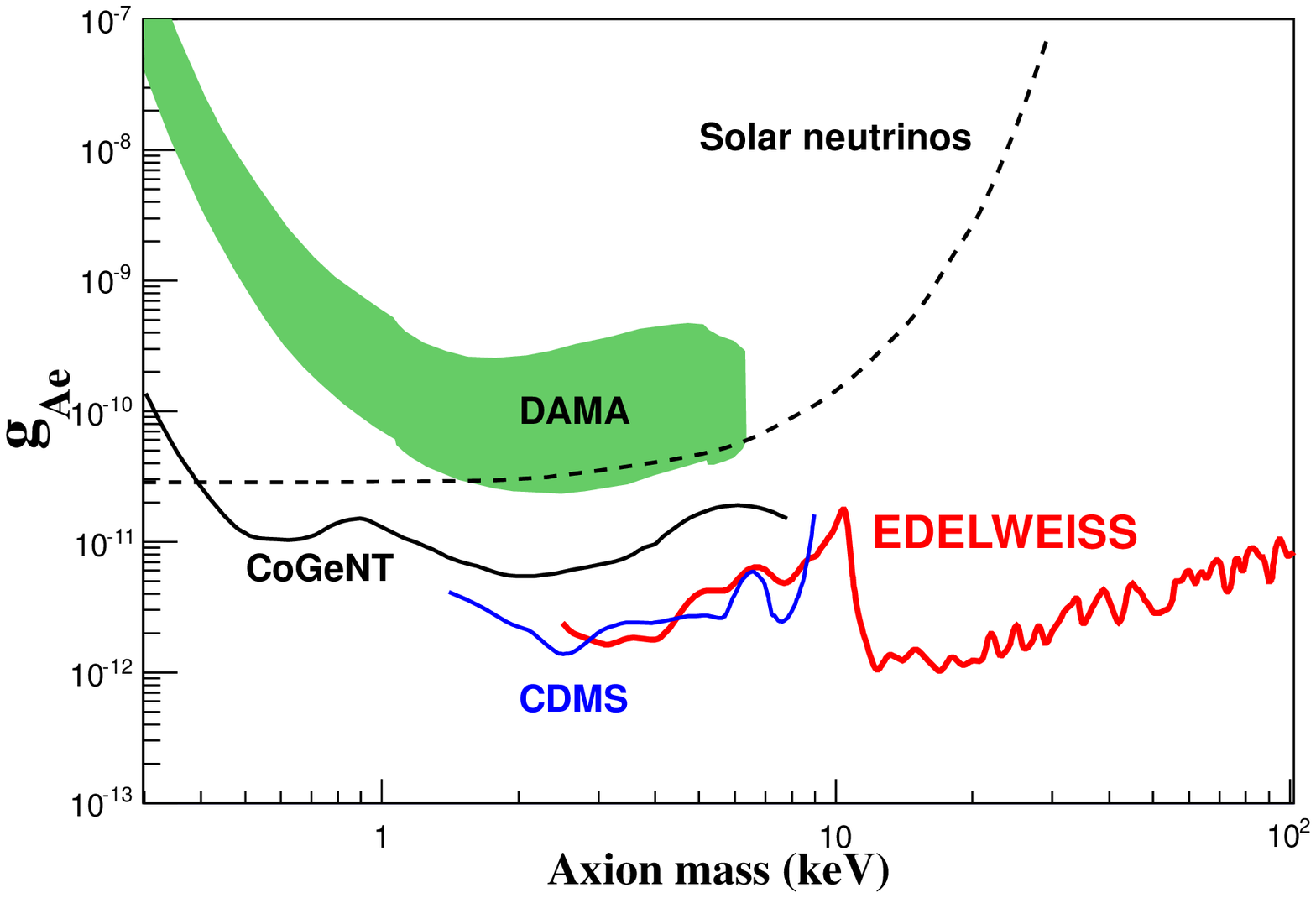}
\caption{Limit on the axion-electron coupling as a function of $m_A$ assuming the local dark matter halo of our galaxy is made entirely of axions. The green contour corresponds to a possible interpretation of the DAMA annual modulation signal~\cite{bib:DAMA_dmaxion}, while the black and blue curves are constraints set by the CoGeNT~\cite{bib:COGENT} and CDMS~\cite{bib:CDMS} germanium detectors respectively. Dashed line: indirect bound derived from the solar neutrino flux measurement~\cite{bib:solar_nu}.}
\label{limit_DM}
\end{figure}

~\\Over the whole energy range, no statistically significant excess was found, except at energies where potential cosmogenic lines are expected. We therefore report a 90\%~CL limit on the axion coupling as a function of its mass. We use the same prescription as above to derive the appropriate limit. The shape of our electron recoil background implies that the strongest constraint is found for 12-~keV axions, for which we found $R_{\text{DM}}<0.05$ counts/kg/d. This translates to a constraint on the dark matter axio-electric coupling at this mass: $g_{ae}<1.05\times 10^{-12}$. The limit on $g_{\rm Ae}$ within this scenario is represented as a function of $m_A$ in Fig.~\ref{limit_DM}.

\section{Results and discussion}
\label{results}

\begin{table}[!ht]
\begin{center}
\begin{tabular}{|c|c|c|c|c|} 
  \hline
  Channel &  14.4 ($g_{Ae}\times g_{AN}^{\rm eff}$) & DM ($g_{Ae}$) & C-B-RD ($g_{Ae}$) & P ($g_{A\gamma}$)  \\
  \hline
 Limit  & $<4.70\times 10^{-17}$ & $<1.05\times 10^{-12}$ & $<2.56\times 10^{-11} $ & $<2.13\times 10^{-9}$~GeV$^{-1}$\\
  \hline
\end{tabular}
\end{center}
\caption{Summary of the limits on the different axion couplings. 14.4 stands for 14.4 keV solar axions, DM for dark matter axions, C-B-RD for Compton-bremsstrahlung and axio-RD axions, and P for Primakoff axions. The quoted values are in the limit $m_A=0$, except for the dark matter case, which is given for $m_A=12.5$~keV. All limits are at 90\% CL except P (95\% CL).}
\label{table_limits}
\end{table}

Using the Sun as a potential source for axions, or under the hypothesis that galactic dark matter is made of axions, we set model-independent constraints on the couplings of ALPs or axions to gamma-rays, electrons, and nucleons. Table~\ref{table_limits} summarizes the limits obtained for each channel on the respective couplings:
\begin{itemize}
	\item	The Primakoff axion search is characterized by a specific time and spectral dependence of the signal. The combination of low effective background and large exposure results in a 95~\% CL limit of $g_{A\gamma}<2.13\times 10^{-9}$~GeV$^{-1}$, an improvement with respect to other germanium crystal-based searches~\cite{bib:CDMS, bib:SOLAX, bib:COSME, bib:DAMA_gay}. Especially for $\sim 1-100$~eV axions, the only other non-crystal constraints are indirect constraints derived from stellar physics. 
	
	\item	In a search for solar axions produced by Compton-bremsstrahlung and axio-RD processes, we provide a model-independent limit on $g_{Ae}$ which is currently the best direct constraint on this coupling compared to~\cite{Derbin_CB, XMASS}. This limit also slightly improves the indirect bound obtained from the solar neutrino flux measurement~\cite{bib:solar_nu}.
	
	\item The search for solar axions emitted by $^{57}$Fe provides first a model-independent limit on  $g_{Ae}\times g_{AN}^{\rm eff}$, which improves the one presented in ~\cite{bib:cuore_RD}. It is possible, as a second approach, to calculate the allowed range of $g_{Ae}$ as a function of the axion mass $m_A$, using values predicted by the DFSZ and KSVZ models for the couplings $g_{AN}^0$ and $g_{AN}^3$.  The resulting constraints are shown in Figure~\ref{14keV_limit}.
		
	\begin{figure}[!ht]
                        \centering
                        \includegraphics[scale=.9]{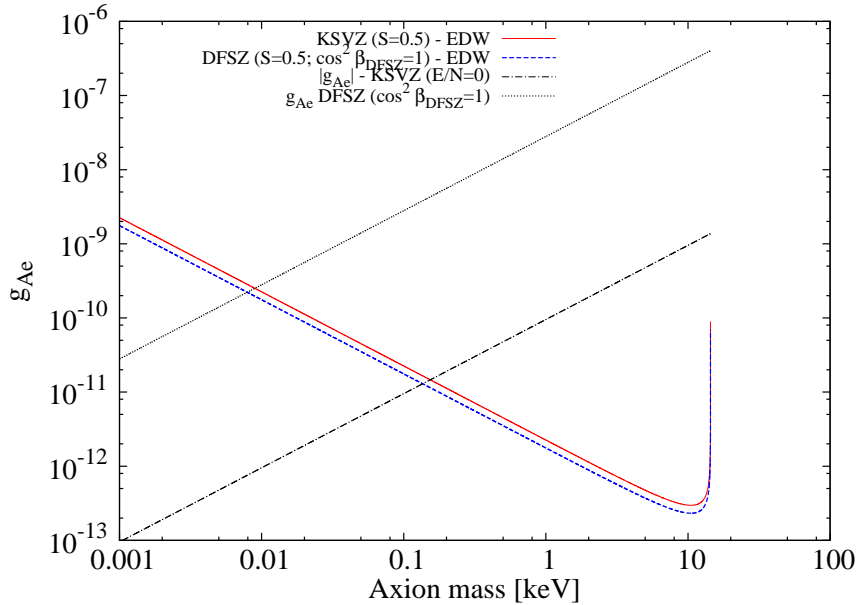}
                        \caption{Limits on $g_{Ae}$ assuming values predicted by the DFSZ and KSVZ models for the couplings $g_{AN}^0$ and $g_{AN}^3$. The curves are calculated with the assumption $S=0.5$ for the flavor-singlet axial vector matrix element in both models and  $\cos^2 \beta _{\rm DFSZ}=1$ for the DFSZ model.}
                     \label{14keV_limit}
\end{figure}

	\item We also tested the specific scenario in which the galactic dark matter halo is made of ALPs with a keV-scale mass. We found a limit similar to the one of CDMS~\cite{bib:CDMS} that we extended up to 100~keV. The limit, as CDMS and CoGeNT \cite{bib:COGENT}, constrains such an interpretation of the observed DAMA feature~\cite{bib:DAMA_dmaxion}.
	
	
\end{itemize}

\begin{figure}[!ht]
\centering
\includegraphics[width=\textwidth]{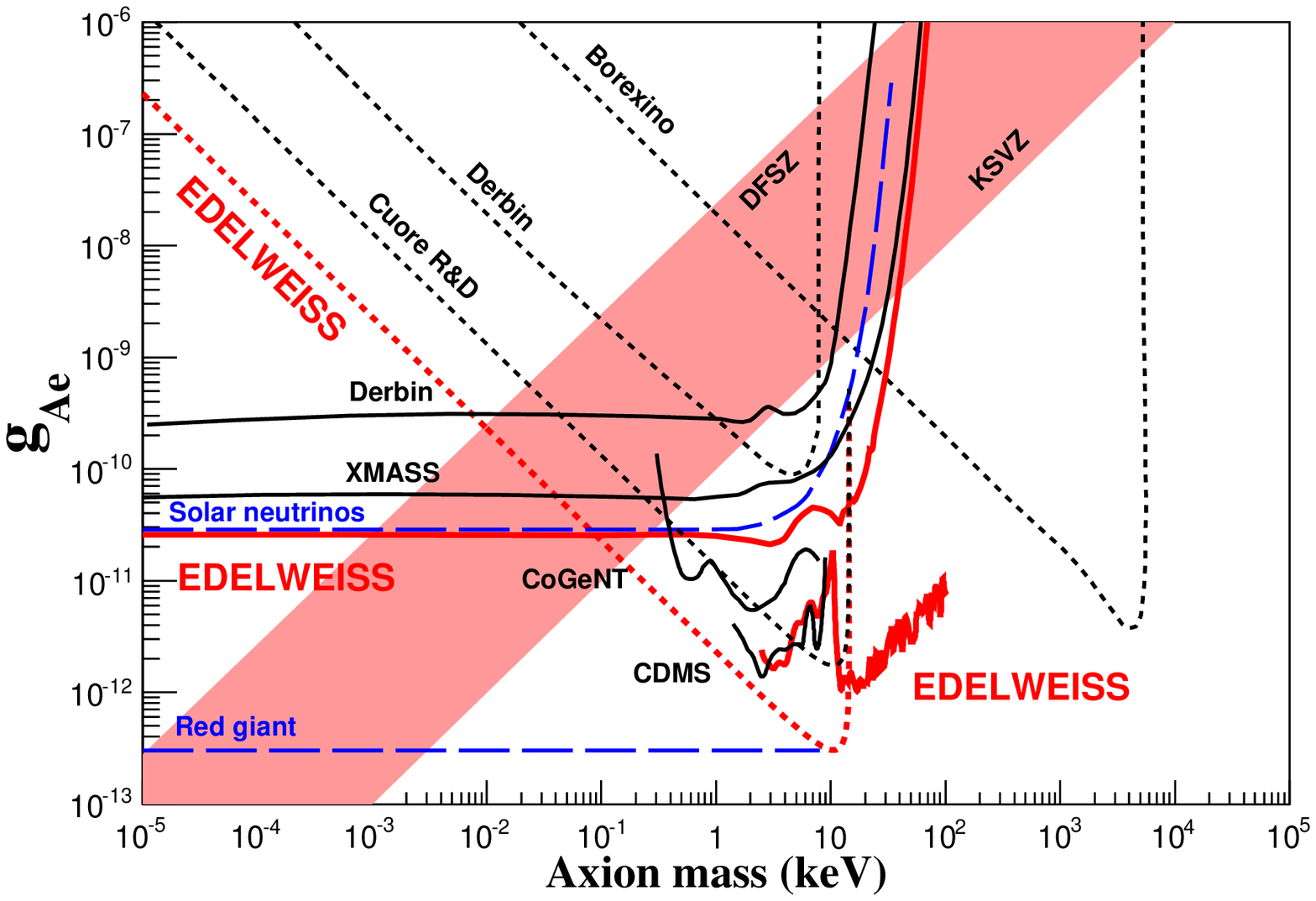}
\caption{Summary of the constraints obtained by EDELWEISS-II on the $g_{Ae}$ axion coupling as a function of $m_A$. The EDELWEISS limits are in red. Continuous lines starting at $m_A=0$, including Derbin~\cite{Derbin_CB} and XMASS~\cite{XMASS}: model-independent limit on $g_{Ae}$ from the solar Compton-bremsstrahlung-recombination flux. Continuous lines in the keV mass range, including CoGeNT~\cite{bib:COGENT} and CDMS~\cite{bib:CDMS}: limits on the coupling of  axions assuming they constitute all local galactic dark matter. Dotted lines, including CUORE R\&D~\cite{bib:cuore_RD}, Derbin~\cite{bib:derbin_Tm} and Borexino~\cite{bib:borex-axions}: bounds on $g_{Ae}$, derived from constraints on $g_{Ae}\times g_{AN}^{\rm eff}$ on various nucleus by assuming that $g_{AN}^{\rm eff}$ follows the DFSZ model with $\cos \beta_{\rm DFSZ}=1$. Benchmark DFSZ and KSVZ models are represented by a shaded band. Indirect astrophysical bounds from solar neutrinos~\cite{bib:solar_nu} and red giants~\cite{bib:red_giants} are represented in dashed lines.}
\label{limit_gae}
\end{figure}

Fig.~\ref{limit_gae} presents a summary of all limits on g$_{Ae}$, some of which are valid only within a given scenario.
	
\begin{table}[!ht]
\begin{center}
\begin{tabular}{|c|c|c|c|c|} 
  \hline
  Channel &  14.4 ($g_{Ae}\times g_{AN}^{\rm eff}$) & C-B-RD ($g_{Ae}$) & P ($g_{A\gamma}$)  \\
  \hline
 KSVZ & $154\,{\rm eV} < m_A < 14.4$~keV& $ 269\,{\rm eV} <m_A < 40$~keV& $ 5.73< m_A\lesssim200$~eV  \\
  \hline
 DFSZ & $7.93\,{\rm eV} < m_A < 14.4$~keV& $ 0.91\,{\rm eV} <m_A < 80$~keV& $ 14.86 < m_A\lesssim200$~eV  \\
  \hline
\end{tabular}
\end{center}
\caption{Excluded ranges of  axion masses derived from EDELWEISS-II constraints within two benchmark models, KSVZ and DFSZ. We assume axion and hadronic parameters described in Section~\ref{sec:sources}. The channels considered are solar 14.4~keV axions (14.4), solar Compton-bremsstrahlung-RD axions (C-B-RD) and solar Primakoff axions (P).}
\label{table_limit_mass}
\end{table}

Within the framework of a given axion model, KSVZ or DFSZ, the only free parameter is the axion mass, or equivalently the Peccei-Quinn symmetry-breaking scale $f_A$ (see Eq.~(\ref{eq:eq2})). Therefore, our limits on the couplings constrain  $m_A$ directly. We calculate the exclusion range for $m_A$ from each of the three solar axion channels previously studied, within both models assuming the model-dependent couplings and hadronic physics parameters given in Section~\ref{sec:sources}. Note that we do not use the dark matter channel here since keV-scale KSVZ or DFSZ axions cannot constitute the local dark matter halo as they are not cold dark matter. In addition, we observe that the limit related to the constraint on $g_{A\gamma}$ holds for mildly relativistic solar axions, up to $m_A \lesssim 200$~eV.
In Table~\ref{table_limit_mass}, the derived limits on the axion mass are summarized for both benchmark models. Within the DFSZ model we completely exclude the mass range $0.91\,{\rm eV}<m_A<80$~keV from the Compton-bremsstrahlung-RD channel only. Other channels are complementary in specific subintervals. Within the KSVZ framework, combining the 14.4 keV and Compton-bremsstrahlung-RD channels, we exclude the mass range $154\,{\rm eV} < m_A < 40$ keV. The Primakoff channel also excludes relativistic solar axions with $m_A>5.73$ eV, therefore closing the window for $5.73\,{\rm eV}<m_A<40$~keV axion masses.




\section{Conclusions}
We have used data from the EDELWEISS-II detectors, originally used for WIMP searches, to constrain the couplings of axion-like particles within different scenarios. Contrary to WIMP searches, we used fiducial \textit{electron} recoils as a potential axion signal. The collected data provide both a reasonably large exposure of up to 448~kg.d, and a good energy resolution of 0.8~keV FWHM at low energy. In addition, the remarkable rejection of near-surface events provided by the ID design, primarily used to reject surface beta radioactivity in WIMP searches, also allows the rejection of additional low-energy gamma-rays of external origin. This rejection provides an average background as low as 0.3 counts/kg/d/keV at 12~keV.

We set new limits on ALP parameters for different scenarios, some of which provide the best bounds for direct axion searches. The 95\% CL bound $g_{A\gamma}<2.13$~GeV$^{-1}$ derived from the solar Primakoff channel constrains axion models in the mass range $\sim 1- 100$~eV for hadronic axions. This constraint is complementary to helioscope bounds, which can currently only probe lower axion masses within these models. Remarkably, the model-independent bound on $g_{Ae}$ obtained from the search for solar Compton-bremsstrahlung-RD axions reaches a better sensitivity than the more indirect bound derived from solar neutrino flux measurements~\cite{bib:solar_nu}.
Combining the results from all solar axion channels provides a wide model-dependent mass exclusion range, $0.91\,{\rm eV}<m_A<80$~keV within the DFSZ framework and $5.73\,{\rm eV}<m_A<40$~keV for KSVZ axions. This is a prominent result for a direct axion search from a single dataset.

We therefore demonstrated the potential of germanium bolometric detectors equipped with the ID design for future ALP searches. Improvements are expected with future setups, such as EDELWEISS-III and EURECA~\cite{bib:EURECA}, thanks to both better energy resolution and larger exposures. 


\acknowledgments
The help of the technical staff of the Laboratoire Souterrain de Modane 
and the participant laboratories is gratefully acknowledged. 
The EDELWEISS project is supported in part by 
the French Agence Nationale pour la Recherche (contract ANR-10-BLAN-0422-03)
and P2IO Labex (Postdoc call 2012), 
the German ministry of science and education (BMBF) within the Verbundforschung
Astroteilchenphysik (grant 05A11VK2),
the Helmholtz Alliance for Astroparticle Physics (HAP) funded by the Initiative
and Networking Fund of the Helmholtz Association,
the Russian Foundation for Basic Research (Russia)
and the Science and Technology Facilities Council (UK). 
We would like to thank also E. Ferrer-Ribas, I. G. Irastorza, B. Laki\'c and T. Papaevangelou for fruitful discussions.



\end{document}